\documentclass[twocolumn,showpacs,preprintnumbers,amsmath,amssymb]{revtex4}

\usepackage{graphicx}

\begin{document}
\preprint{APS/123-QED}
\title{Intrinsic Spin and Orbital-Angular-Momentum Hall Effect}

\author{S. Zhang and Z. Yang}
\affiliation{Department of Physics and Astronomy, University of
Missouri-Columbia, Columbia, MO 65211}
\date{\today}

\begin{abstract}
A generalized definition of intrinsic and extrinsic
transport coefficients is introduced. 
We show that transport coefficients from the intrinsic origin are
solely determined by local electronic structure, and thus the
intrinsic spin Hall effect is not a transport phenomenon. 
The intrinsic spin Hall current is always accompanied by an equal but 
opposite intrinsic orbital-angular-momentum Hall current.
We prove that the intrinsic spin Hall effect 
does not induce a spin accumulation at the edge of the sample or near the
interface.

\end{abstract}


\maketitle

Recently, there are emerging {\em theoretical} interests on the
spin Hall effect in a spin-orbit coupled system 
\cite{Sinova,Culcer,Nomura,Murakami,Rashba,Rashba2,Loss,Loss2,
Loss3,Inoue,Mishchenko,Dimitrova,Hu,Ma}. The spin Hall effect
refers to a non-zero {\em spin} current in the direction transverse to
the direction of the applied electric field. Earlier 
studies had been focused on an extrinsic effect \cite{Hirsch,Zhang}, 
namely, when conduction
electrons scatter off an impurity with the spin-orbit interaction, 
the electrons tend to deflect to the left (right) more than to the
right (left) for a given spin orientation of the electrons. Thus the
impurity is the prerequisite in the extrinsic spin Hall effect. Recently, 
the spin Hall effect has been extended to semiconductor heterostructures
where the spin-orbit coupled {\em bands} are important. 
It has been shown that the spin current exists in the absence of impurities,
termed as the intrinsic or dissipationless spin Hall effect (ISHE) in order
to distinguish
the impurity-driven extrinsic spin Hall effect (ESHE) mentioned above. 
In general, the magnitude of ISHE is two to three orders larger than
that of ESHE; this immediately generates an explosive interest in
theoretical research on the ISHE since
the spin current is regarded as
one of the key variables in spintronics application.

However, the spin current generated via ISHE is fundamentally different from 
conventional spin-polarized transport in many ways. First,
the spin current is carried by the entire spin-orbit coupled Fermi sea, 
not just electrons or holes at the Fermi level \cite{Murakami,Sinova}. Second, 
ISHE exists even for an equilibrium system (without external electric 
fields) \cite{Rashba} 
and ISHE is closely related to the dielectric response function that 
characterizes the electronic deformation \cite{Rashba2}. Most recently, it is
proposed that the intrinsic spin Hall effect exists even in 
insulators \cite{Murakami2}. The above unconventional
properties cast serious doubts on experimental relevance  
of the intrinsic spin current. It has been already alerted by Rashba 
\cite{Rashba,Rashba2} that
the ISHE may not be a transport phenomenon. Due to the ill-defined 
nature of the spin current in the spin-orbit coupled Hamiltonian,  
theories utilizing different approaches produce contradicting results:
some predicted a zero spin Hall current in the presence of an arbitrary weak
disorder and some claimed a universal spin conductivity at weak disorder.
In this letter, we do not try to resolve the above theoretical debate,
instead we reveal the spurious nature of the intrinsic spin Hall effect and
discuss its experimental consequences. We first define generalized 
intrinsic and extrinsic transport coefficients from the semiclassical 
transport equation. We show that the intrinsic spin current
is always accompanied by an equal but opposite orbital-angular-momentum (OAM)
current for a spin orbit coupled system. Thus, the 
intrinsic magnetization current which is the sum of the spin current and the
orbital angular momentum current is identically zero. Next, we construct the
equation of motion for the spin density in the presence of the intrinsic
and extrinsic mechanisms. We find that the intrinsic spin current is exactly
canceled by a spin torque and thus the spin accumulation at the edge of the
sample is solely determined by the extrinsic spin current. 
The above results make us conclude that the intrinsic spin current has 
{\em no experimental consequences} in terms
of the spin transport measurement for an arbitrary strength of the 
intrinsic Hall conductivity.
Therefore, the intrinsic spin current is a pure theoretical object, at least,
in the limit of the semiclassical picture of the spin transport.
Finally, we brief comment on the most recent experimental results
\cite{Wunderlich}.
 
Let us consider a spin-dependent Hamiltonian
\begin{equation}
H= - \frac{\hbar^2}{2m} \nabla^2 + V({\bf r}, \mbox{\boldmath $\sigma$})
+ eEx + V_i ({\bf r}, \mbox{\boldmath $\sigma$})
\end{equation}
where the second term represents a periodic spin-orbit potential, the
third term is the interaction with a DC electric field $E$ in the 
$x$-direction, and the last term
is the impurity potential that may or may not depend on spin.
Now let us consider how an arbitrary dynamic variable $\hat{G}$ responds
to the electric field. A standard  semiclassical version
of the linear response function ${\cal J} ({\bf r})$ is
\begin{equation}
{\cal J}({\bf r}) = \sum_{{\bf k} \lambda} G_{{\bf k}\lambda}
f( \epsilon_{{\bf k}\lambda} , {\bf r} ) 
\end{equation}
where $\hat{G}$ can be any dynamic variable such as the current
density, the spin current density, the magnetic moment, etc.,
$G_{{\bf k}\lambda}  = \int d{\bf r} \Psi^+_{{\bf k}\lambda} (\bf r)
\hat{G} \Psi_{{\bf k}\lambda} (\bf r) \equiv
<\Psi_{{\bf k}\lambda}|\hat{G}| \Psi_{{\bf k}\lambda}>$ is the expectation 
value for the eigenstate $\Psi_{{\bf k}\lambda} (\bf r)$ (Bloch states) 
determined by the first three terms in Eq.~(1), $\lambda = \pm 1 $ represents
the index of the spin sub-band, and $f( \epsilon_{{\bf k}\lambda} ,
{\bf r} )$ is the distribution function that depends on the detail of
the scattering potential $V_i ({\bf r}, \sigma)$, the last term of Eq.~(1).
The dependence of ${\cal J} ({\bf r})$ on the electric field enters in 
two places:
the wavefunctions and the distribution function. We may expand them up to the
first order in the electric field. The wavefunction is written as,
\begin{equation}
\Psi_{{\bf k}\lambda} ({\bf r}) = \Psi^{(0)}_{{\bf k}\lambda}
({\bf r}) + \Psi^{(1)}_{{\bf k}\lambda} ({\bf r})
\end{equation}
where $\Psi^{(0)}_{{\bf k}\lambda}({\bf r})$ is the
unperturbed electronic structure determined by the first two
terms in Eq.~(1), and 
\begin{equation}
\Psi^{(1)}_{{\bf k}\lambda} ({\bf r}) = \sum_{{\bf k}' \lambda ' \neq
{\bf k}\lambda }
\frac{<\Psi^{(0)}_{{\bf k}'\lambda'} | eEx | \Psi^{(0)}_{{\bf k} \lambda}>}
{\epsilon_{{\bf k}\lambda} - \epsilon_{{\bf k}'\lambda'}} 
\Psi^{(0)}_{{\bf k}'\lambda'} ({\bf r}), 
\end{equation}
is the first order perturbation to the third term in Eq.~(1). 
Similarly, we write the distribution function in terms of 
the equilibrium and non-equilibrium parts,
\begin{equation}
f( \epsilon_{{\bf k}\lambda} , {\bf r} ) = f^0
( \epsilon_{{\bf k}\lambda} ) + \left( - \frac{\partial f^0}{\partial 
\epsilon_{{\bf k}\lambda}} \right)
g({\bf k}\lambda, {\bf r}) 
\end{equation}
where $f^0$ is the equilibrium distribution function and the non-equilibrium
function $g({\bf k}\lambda, {\bf r})$ is proportional to the electric field.
By placing Eqs.~(3) and (5) into Eq.~(2) and keeping only the first order
term in the electric field, we have ${\cal J}({\bf r})
\equiv {\cal J}_{int}+{\cal J}_{ext}$ where 
\begin{equation}
{\cal J}_{int} = 2 {\rm Re} \sum_{{\bf k} \lambda}
<\Psi^{(0)}_{{\bf k}\lambda}
|\hat{G}|\Psi^{(1)}_{{\bf k}\lambda}> f^0 (\epsilon_{{\bf k}\lambda})
\end{equation}
is defined as the intrinsic linear response and ${\rm Re}$ stands for
the real part, and 
\begin{equation}
{\cal J}_{ext} = \sum_{{\bf k} \lambda} <\Psi^{(0)}_{{\bf k}\lambda}|\hat{G}|
\Psi^{(0)}_{{\bf k}\lambda}>
\left( - \frac{\partial f^0}{\partial
\epsilon_{{\bf k}\lambda}} \right)
g({\bf k}\lambda, {\bf r})
\end{equation}
is called the extrinsic linear response. The above distinction between
intrinsic and extrinsic contributions to the transport properties has been
already introduced by a number of groups, in particular, by Jungwirth
{\em et al.} \cite{Jungwirth} in their study of the anomalous Hall effect in 
itinerant ferromagnets. Equation (6) shows that the
intrinsic linear response coefficient is not related to the transport
phenomenon since ${\cal J}_{int}$ is determined by the equilibrium 
distribution function and the {\em local electronic structure}. 
Thus, there are no transport
length scales such as the mean free path or spin diffusion length 
in ${\cal J}_{int}$. The extrinsic linear response,
${\cal J}_{ext}$, is a true transport quantity because it is directly 
proportional to the non-equilibrium distribution function that is determined 
by various scattering mechanisms. At low temperature, 
the factor of $\partial f^0/\partial \epsilon_{{\bf k}\lambda}$ limits 
the transport states to the Fermi level. Comparing Eq.~(6) and Eq.~(7),
we realize that the intrinsic effect is simple and easy to calculate while 
the extrinsic effect is much more complicated. As long as we know the
Bloch states $\Psi^{(0)}_{{\bf k}\lambda}$, the intrinsic transport 
coefficient can be straight-forwardly evaluated since $f^0$ is known. 
The extrinsic transport
coefficient not only depends on the Bloch states, but also on
the non-equilibrium distribution function that is usually the center of
the relevant physics. Here, however, we should concentrate on the
easy problem: calculation of the intrinsic transport from Eq.~(6) by using
a model Hamiltonian. 

We choose a Rashba Hamiltonian to illustrate the physics of the
intrinsic transport properties. A similar calculation can also be
performed for a Luttinger Hamiltonian \cite{Luttinger}.
For the Rashba Hamiltonian, the second term of Eq.~(1) is
$V({\bf r}, \sigma) = (\alpha/\hbar) 
\mbox{\boldmath $\sigma$} \cdot ({\bf p} \times \hat{\bf z}) $
where {\bf p} is the momentum in $xy$ plane, $\mbox{\boldmath $\sigma$}$
is the Pauli matrix and $\alpha$ is the coupling constant. 
Before we calculate the spin and the OAM 
Hall currents from Eq.~(6), we list the wavefunction and
the dispersion relation of the Rashba Hamiltonian so that one can easily follow
our derivation at each step
\begin{equation}
\Psi^{(0)}_{{\bf k}\lambda} ({\bf r}) = \frac{e^{i{\bf k}\cdot {\bf r}}}
{\sqrt{2A}} \left( \begin{array}{c}
1 \\
-i \lambda (k_x +ik_y)k^{-1} \\
\end{array}
\right) 
\end{equation}
where $A$ is the area of the 2-dimensional electron gas, and 
\begin{equation}
\epsilon_{{\bf k}\lambda} = \frac{\hbar^2 k^2}{2m} + \lambda \alpha k
\end{equation}
where $k=|{\bf k}|=\sqrt{k_x^2 + k_y^2}$.
By placing above two equations into (4), we have \cite{Rashba}
\begin{equation}
\Psi^{(1)}_{{\bf k}\lambda} ({\bf r}) = - \frac{\lambda eE k_y}{
4 \alpha k^3}
\Psi^{(0)}_{{\bf k}-\lambda} ({\bf r}). 
\end{equation}
In obtaining the above result, we have used $<\Psi^{(0)}_{{\bf k}'\lambda'} 
|x|\Psi^{(0)}_{{\bf k}\lambda}> = -\frac{\lambda \lambda'}{2} \delta_{
{\bf kk}'} \frac{k_y}{k^2} $.

We now proceed to calculate the spin Hall current by taking the operator 
$\hat{G}= (1/2)[ s_z v_y + v_y s_z] $
where $s_z = (\hbar/2)\sigma_z$ is the z-component of the spin operator and
$v_y$ is the y-component of the velocity. By placing the above definition
along with Eqs.~(8) and (10), and by taking the distribution function
a step function at zero temperature, we obtain the intrinsic spin Hall current
from Eq.~(6),
\begin{equation}
{\cal J}^{spin}_{int} = \frac{e}{8\pi} E;
\end{equation}
where we have assumed that the Fermi energy is larger than the
spin-orbit coupling energy so that both spin sub-bands cross the Fermi level.
Equation (11) represents the universal spin conductivity ($e/8\pi$) obtained by
many groups \cite{Murakami,Sinova}. 

Our central question is: what is the physical meaning
of this spin current derived from the equilibrium distribution function? 
To see clearly what this spin current represents, we recall that the spin
is not a conserved quantity in a spin-orbit coupled system. If one is 
interested in the {\em magnetization current} or the total angular
momentum current, one should also include the OAM Hall current. 
The OAM Hall current can be similarly calculated 
by introducing an operator for the OAM Hall current 
\begin{equation}
\hat{G} =\frac{1}{2} \left[
({\bf r}\times {\bf p})_z v_y + v_y ({\bf r}\times {\bf p})_z
\right]
\end{equation} 
where $({\bf r}\times {\bf p})_z = xp_y-yp_x$ is $z$-component of the
OAM. The same straightforward evaluation 
of Eq.~(6) leads to 
\begin{equation}
J^{orbit}_{int} = - \frac{e}{8\pi} E.
\end{equation} 
Thus the OAM current is exactly equal and opposite 
to the spin current. This result is not surprising at all: 
the total angular momentum ($z$-component), spin plus orbital, is 
conserved for the Rashba Hamiltonian and thus we can choose the Bloch 
states that are simultaneous eigenstates of the total angular momentum
and the Hamiltonian. In fact, one can directly show that the total angular 
momentum current vanishes if we use $[s_z +L_z, H] =0$ to the Rashba
Hamiltonian. 

Having discussed that the intrinsic spin current is always accompanied with
the OAM current in a bulk spin-orbit coupled material, our next question
is whether the intrinsic spin current can produce a spin accumulation at the
edge of the sample? To answer this question, we recall the basic idea
of the spin accumulation for the extrinsic spin current. 
When an extrinsic spin current spatially varies, non-equilibrium spins 
will be accumulated so that the spin-diffusion is balanced by the spin-drift
current. Equivalently, the spin accumulation results in the chemical 
potential splitting between two
spin sub-bands and a voltage can be measured experimentally when the sample
is attached to a ferromagnetic lead \cite{Zhang}. Mathematically,  
the non-equilibrium spin-dependent chemical potential or spin accumulation
is the average of the {\em non-equilibrium distribution function} 
\cite{Valet}. For 
the intrinsic spin Hall current, the distribution is an equilibrium
distribution and one would expect that the concept of the
spin-dependent chemical potential breaks down. Indeed, we show next that the
the intrinsic spin Hall current does not lead to spin accumulations at the
sample edge and across an interface.

To calculate the spin accumulation or the position-dependent spin density
${\bf S} ({\bf r}, t)$ at the edge of the sample or across an interface,
one relies on the semiclassical equation of motion that can be generally
written as
\begin{equation}
\frac{\partial {\bf S} ({\bf r}, t)}{\partial t} + \mbox{\boldmath $\nabla$}
\cdot [ {\bf J}_{int} + {\bf J}_{ext}] = \mbox{\boldmath $\tau$}
+ \mbox{\boldmath $\tau$}_{ext} + \left( \frac{\partial {\bf S}}{\partial t}
\right)_{colli} 
\end{equation}
where ${\bf J}_{int}$ and ${\bf J}_{ext}$ are the intrinsic and extrinsic
spin current densities, $\mbox{\boldmath $\tau$}$ and
$\mbox{\boldmath $\tau$}_{ext}$ are the intrinsic and extrinsic spin torques
due to non-commutivity of the Hamiltonian with the spin operator, i.e., the
spin torque is calculated by replacing $\hat{G}$ by $[{\bf s}, H]/i\hbar$ 
in Eqs.~(6) and (7),
and the last term in  Eq.~(14) is a collision term that is to relax the
nonequilibrium distribution function to an equilibrium one. 
To explicitly obtain the spin accumulation ${\bf S} ({\bf r}, t)$ in
a closed form, it is necessary to use a wave-package description so that
the position-dependence can be readily included. Culcer {\em et al.} 
\cite{Culcer} have already formulated that the spin torque
can be written as two terms. In our notation, we find
the expression for the intrinsic spin torque is
$\mbox{\boldmath $\tau$} = \mbox{\boldmath $\tau$}_{0}+
\mbox{\boldmath $\tau$}_1 $
where
\begin{equation}
\mbox{\boldmath $\tau$}_0 = \sum_{{\bf k}\lambda} <{\bf k}\lambda|
\frac{1}{i\hbar}[{\bf s},H]|
{\bf k}\lambda> f^0 ({\bf k}\lambda, {\bf r} )
\end{equation}
and 
\begin{equation}
\mbox{\boldmath $\tau$}_1 = - \mbox{\boldmath $\nabla$}\cdot 
\sum_{{\bf k}\lambda}
<{\bf k}\lambda|\frac{1}{i\hbar}[{\bf s},H] {\bf r} | {\bf k}\lambda>
f^0 ({\bf k}\lambda, {\bf r})
\end{equation}
where the symmetrization of the product of $[{\bf s},H]$ and ${\bf r}$ is
implied. We emphasize that $\mbox{\boldmath $\tau$}_1$ comes from the
position-dependence of the center of the wave-packet.
By using the fact that the wavefunction is an eigenstate 
of the Hamiltonian, $H|{\bf k}\lambda> = E_{{\bf k}\lambda}
|{\bf k}\lambda>$, we immediately see that the expectation value of 
$[{\bf s}, H]$ is zero, i.e., $\mbox{\boldmath $\tau$}_0 = 0$. To calculate
$\mbox{\boldmath $\tau$}_1$, we use 
the commuting relation $[H,{\bf r}] = -i \hbar {\bf v} + 
i \alpha ({\bf e}_z \times 
\mbox{\boldmath $\sigma$})$, where ${\bf v}$ is the velocity operator. 
After a straight-forward algebra simplification, we have found \cite{footnote} 
\begin{equation}
\mbox{\boldmath $\tau$} = \mbox{\boldmath $\nabla$} \cdot 
\sum_{{\bf k}\lambda} <{\bf k}\lambda|\frac{{\bf vs}+{\bf sv}}{2}
| {\bf k}\lambda> f^0 ({\bf k}\lambda, {\bf r} )
\equiv \mbox{\boldmath $\nabla$}\cdot {\bf J}_{int} .
\end{equation}
Therefore, the intrinsic spin torque exactly equals the divergence of the 
intrinsic spin current. The equation of motion, Eq.~(14), now becomes
\begin{equation}
\frac{\partial {\bf S} ({\bf r}, t)}{\partial t} + \mbox{\boldmath $\nabla$}
\cdot {\bf J}_{ext} = \mbox{\boldmath $\tau$}_{ext}
+ \left( \frac{\partial {\bf S}}{\partial t} \right)_{colli}.
\end{equation}
We conclude that the intrinsic spin current does not enter into the play
in the equation of motion. The spin accumulation 
${\bf S} ({\bf r}, t)$ 
is solely determined by the extrinsic part of the current density
${\bf j}_{ext}$, the spin torque $\mbox{\boldmath $\tau$}_{ext}$, 
and the spin relaxation in the collision term. 

We now return to our central issue on the problem, namely, whether the
intrinsic spin Hall conductivity can be measured via conventional meanings
of the spin transport. Since the spin current is not directly measurable, 
two schemes are usually employed: one is the realization of measuring
the electric field induced by the {\em magnetization current} \cite{Meier}
and the other is the
spin accumulation at the sample of the edge or across an interface
\cite{Johnson}. We should discuss them separately below.

If there is a net magnetization current in a bulk material, a circular
electric field outside the sample will be induced. This phenomenon is analogous
to the magnetic field induced by a charge current, known as Biot-Savart law or
Ampere's law. For example, it was proposed that a spin current generated
via spin waves propagation through a nanowire can be detected
by an induced electric field just outside the nanowire
\cite{Meier,Hong}. However, the above 
proposal is applied to the case where the OAM current
is absent. In the present case, the OAM current
is exactly opposite to the spin current so that the net magnetization current
is zero. Therefore, we conclude that there is no electric field associated
with the intrinsic spin current.

The more efficient method to detect the spin current is by measuring the
spin accumulation due to spatial variation of the spin current, e.g., 
the Johnson-Silsbee's experiment \cite{Johnson}. Based on the equation of
motion given by Eq.~(18), the divergent of the {\em extrinsic but not
intrinsic} spin currents can lead to a buildup of spin accumulation. To 
determine the spin accumulation, one usually makes a relaxation-time
approximation so that the collision term in Eq.~(18) is modeled by 
$ -{\bf S}/\tau_{sf}$. Since the intrinsic spin current does not
contribute to the equation of motion for the spin accumulation, the
measurement based on the detection of the spin accumulation will produce
a null contribution from the spin Hall effect, no matter how large the
intrinsic spin current is. 

Two experimental groups have recently observed the spin Hall effect
by detecting spin accumulation at the edges of the samples \cite{Wunderlich}. 
Kato {\em et al.} argued that the effect is extrinsic based on their
experimental results that the spin accumulation is independent of the strain
direction, while Wunderlich {\em et al} claimed that their observed effect is
intrinsic based on the assumption that the impurity scattering is weaker
than the spin-orbit coupling in their samples (clean limit).
We point our here that the clean limit in the experiment 
does not imply the spin accumulation from the intrinsic origin.
Instead, our analysis has shown that no matter how large is the intrinsic
spin current, the observed effect has to be an extrinsic origin because the
spin accumulation is independent of the intrinsic spin current.

We finally draw a picture on why the intrinsic spin Hall fails to
produce experimental consequences. Consider a contact between a 
Rashba material and a
non-magnetic material with no spin-orbit coupling. The spin current,
as well as the orbital angular momentum current, would exist in the
the Rashba material. However, the spin current drops
to zero across the interface of the non-magnetic material, i.e., the spin
current is not continuous; this is because the spin torque 
produces a mechanism to transfer the spin current to the 
orbital angular momentum current or vice versus. As a result, when the
spin orbit coupling vanishes at the non-spin-orbit coupled material, 
both the spin and orbital angular momentum currents drop to zero. The loss of
the spin current exactly equals to the gain of the OAM 
current so that the total angular momentum current or magnetization
current is continuous across
the interface of the layers; they are both zero. 
For the same reason, the edge of the sample never develops
spin accumulation because the usual boundary condition of zero
spin current at the surface is no more valid, instead, the total
angular momentum current is zero at the surface for the intrinsic spin Hall
effect. 

In conclusion, we have constructed a general framework for calculating
intrinsic linear response coefficients. We have shown that the 
intrinsic spin Hall effect is accompanied by the intrinsic 
orbital-angular-momentum Hall effect
so that the magnetization current is zero in a spin-orbit 
coupled system. The intrinsic spin Hall effect is not a useful source
of spin currents because the intrinsic spin current does not enter into
the equation of motion for the spin transport.
Most of the proposed experimental detections
of the intrinsic spin Hall effect are the artifact of the boundary
conditions that are not valid for the intrinsic spin Hall current.
 
This work is partially supported by DARPA-SPINS and by NSF-DMR-0314456.

\end{document}